\documentclass[nonacm=true,manuscript,screen,format=sigplan]{acmart}
\AtBeginDocument{%
  \providecommand\BibTeX{{%
    \normalfont B\kern-0.5em{\scshape i\kern-0.25em b}\kern-0.8em\TeX}}}

\setcopyright{none}
\copyrightyear{2023}
\acmYear{2023}
\acmDOI{XXXXXXX.XXXXXXX}

\acmConference[Haskell '23]{ACM SIGPLAN Haskell Symposium 2023}{Sep 4--9, 2023}{Seattle, Washington, USA}
\acmBooktitle{Proceedings of the 16th ACM SIGPLAN International Haskell Symposium (Haskell ’23), Sep 4--9, 2023, Seattle, Washington, USA} 
\acmPrice{15.00}
\acmISBN{978-1-4503-XXXX-X/18/06}

\begin{document}

\title{An Auto-Parallelizer for Distributed Computing in Haskell}

\author{Yuxi Long}
\authornote{These authors contributed equally to this research.}
\email{yuxi.long@duke.edu}
\orcid{0000-0002-0833-2910}
\affiliation{%
  \institution{Duke University}
  \streetaddress{308 Research Drive}
  \city{Durham}
  \state{North Carolina}
  \country{USA}
  \postcode{27710}
}

\author{Shiyou Wu}
\authornotemark[1]
\email{shiyou.wu@duke.edu}
\orcid{0000-0002-3346-6611}
\affiliation{%
  \institution{Duke University}
  \streetaddress{308 Research Drive}
  \city{Durham}
  \state{North Carolina}
  \country{USA}
  \postcode{27710}
}

\author{Yingjie Xu}
\authornotemark[1]
\email{yingjie.xu@duke.edu}
\orcid{0009-0001-1201-1968}
\affiliation{%
  \institution{Duke University}
  \streetaddress{308 Research Drive}
  \city{Durham}
  \state{North Carolina}
  \country{USA}
  \postcode{27710}
}

\renewcommand{\shortauthors}{Long, Wu and Xu}





\received{TBD}
\received[revised]{TBD}
\received[accepted]{TBD}

\begin{titlepage}
   \begin{center}
   \vspace*{\fill}
   \huge
   Submitted to the \textit{28$^{\text{th}}$ ACM SIGPLAN International Conference on Functional Programming, Haskell Symposium}.\\
   \vspace{1.5cm}
   Accepted for oral presentation.\\
   Presented Sep 8, 2023.
   \vspace*{\fill}         
   \end{center}
\end{titlepage}

\maketitle

\section{Introduction}

In this paper, we designed, implemented, and benchmarked a Haskell auto-parallelizer with a simple yet powerful interface by taking advantage of the default purity of Haskell functions. One of the main challenges in distributed computing is building interfaces and APIs that allow programmers with limited background in distributed systems to write scalable, performant, and fault-tolerant applications on large clusters. While accounting for the complex side-effects induced by the separate pieces of code running on different computers, designers of these frameworks often have to find the compromise between more limited API or looser consistencies and semantics.  For example, \textit{MapReduce} \cite{62} provides the programming models of \textit{Map} and \textit{Reduce} that completely abstracts the distributed nature of the program away from the programmer, with the caveat that side-effects have to be atomic and idempotent. \textit{Ray}, on the other hand, asks the programmer to decide whether the execution should be stateful or not. In large projects, it could be very difficult for a programmer to reason about these properties because some library deep down in the nested function calls could introduce unexpected side effects, such as writing to the same temporary file across all calls, which would result in incorrect execution of the program.

By design, the purity of Haskell mitigates these problems drastically and allows us to provide a much simpler interface than traditionally possible with other object-oriented languages. Since the purity of a function call can be directly inferred from its type signature at compile time, we can parallelize the pure functions across the workers without worrying about the side-effects while also ensuring that the impure functions are executed in order. More specifically, given a Haskell program and a section of the code to parallelize, a scheduler can parse the program's data dependencies between function calls and greedily schedules tasks to worker nodes as their inputs are ready, all without requiring the user to understand the way in which underlying libraries execute.




We have built a prototype implementing the ideas above. While our prototype does not yet have industrial-scale scalability or fault-tolerance, it has shown promises as a simple way to achieve parallel speed-up on existing workloads. Hence, in this paper, we benchmarked our implementation to illustrate the potential for future work in this direction.


\section{Design} \label{sec:design}


The user specifies which section of the code to parallelize. This should be a section with high levels of abstraction (in other words, each function call takes some amount of time to execute). For example, a user working on natural language processing could write something like the following:

\begin{verbatim}
data Summary = ...  -- A custom data type

clean_files :: IO Summary
clean_files = ...

complex_evaluation :: Summary -> Int
complex_evaluation x = ...

semantic_analysis :: IO Int
semantic_analysis = ...

main :: IO()
main = do
    x <- clean_files
    let y = complex_evaluation x
    z <- semantic_analysis
    print (y, z)
\end{verbatim}

In this case, the user should specify that the \verb|main| function is the function they want to parallelize, because it calls on other high-level functions. From there, a parser could infer the following data dependency graph:

\begin{figure}[htpb]
    \centering
    \includegraphics[width=0.2\textwidth]{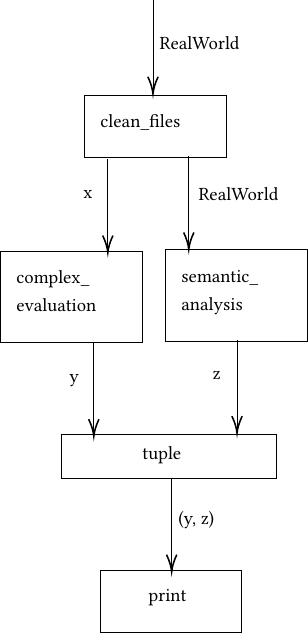}
    \caption{Data dependency graph generated from example Haskell program}
    \label{fig:dep-graph}
\end{figure}

Notice that \verb|RealWorld| is considered an input and output by each \verb|IO| function. In this case, once \verb|clean_files| is done, both \verb|complex_evaluation| and \verb|semantic_analysis| can be scheduled for execution.

Another example is a deep learning project, in which the user specify the forward and backward passes of the neural network. In practice, a user is usually writing high-level code, so the section worth parallelizing usually coincides with the section that the user writes, which is very convenient. In our prototype, only the main function is parallelized, but we could see a future implementation where the user can specify any arbitrary function.

While we only implemented a shallow parser in our implementation, we think that incorporating a more powerful parser such as \href{https://github.com/aaronallen8455/graph-trace}{Graph Trace} that can parse arbitrary depth could further allow the user to specify the granularity of distribution.







\section{Related Works}



This section provides both works that tread similar paths as us and works that a future implementation of this idea could build on.


There are a number of automatic parallelization compilers for other languages. These compilers parallelize at instruction level on a shared memory machine. Many such research compilers exist for Fortran and C, such as the Vienna Fortran Compiler \cite{benkner1999vfc} and the Intel C++ Compiler \cite{intelcppcompiler}. A recent work by Google \cite{zheng2022alpa} enables automatic parallelization of models written under the JAX framework \cite{jax2018github}.

There are a number of existing packages for shared-memory parallel in Haskell. Built-in to Haskell are the \textit{Control.concurrent} package provides basic APIs to threads and forks, and the \textit{parallel} library provides primitives such as \texttt{par} and \texttt{pseq}. Projects such as \href{https://hackage.haskell.org/package/monad-par}{monad-par} \cite{marlow2011monad} and \href{https://hackage.haskell.org/package/lvish}{lvish} \cite{kuper2013lvars} provide additional interfaces through monads to enable work-stealing schedulers.

A few libraries exist as backbones for distributed computing. \href{https://haskell-distributed.github.io/}{Cloud Haskell} \cite{epstein2011towards} provides APIs for serializing functional closures and channels for network communications. The \href{https://hackage.haskell.org/package/network}{\textit{network}} package provides a low-level networking interface. A recent advancement is \href{https://github.com/tweag/sparkle}{Sparkle} \cite{tweag2023}, which provides a Haskell interface to Spark \cite{zaharia2012resilient}, a unified framework for resilient distributed applications written in Scala. We believe future implementations could benefit from Sparkle for its robustness.

Finally, while we wrote a custom script for generating dependency graph based on a shallow parsing of the program in our implementation for simplicity, there are robust libraries that can generate functional graphs for complex Haskell programs. Two that we found are \href{https://github.com/aaronallen8455/graph-trace}{Graph Trace} and \href{https://hackage.haskell.org/package/SourceGraph}{SourceGraph}, which future implementations could benefit from.







\section{Experiments} \label{sec:experiments}

We performed experiments on matrix operations (generation and multiplication of large random matrices) with different numbers of workers simulated using Cloud Haskell. Although this may seem simple, it is the foundation for modern deep learning computations and good performance and scalability would indicate great potential for future generalization. For reference, we used single-thread and Haskell's built-in SMP parallelism as our baselines. The task size is the number of times that the matrix operations are performed, and all results are rounded to the nearest second. 

\begin{figure}[htpb]
    \centering
    \includegraphics[width=0.5\textwidth]{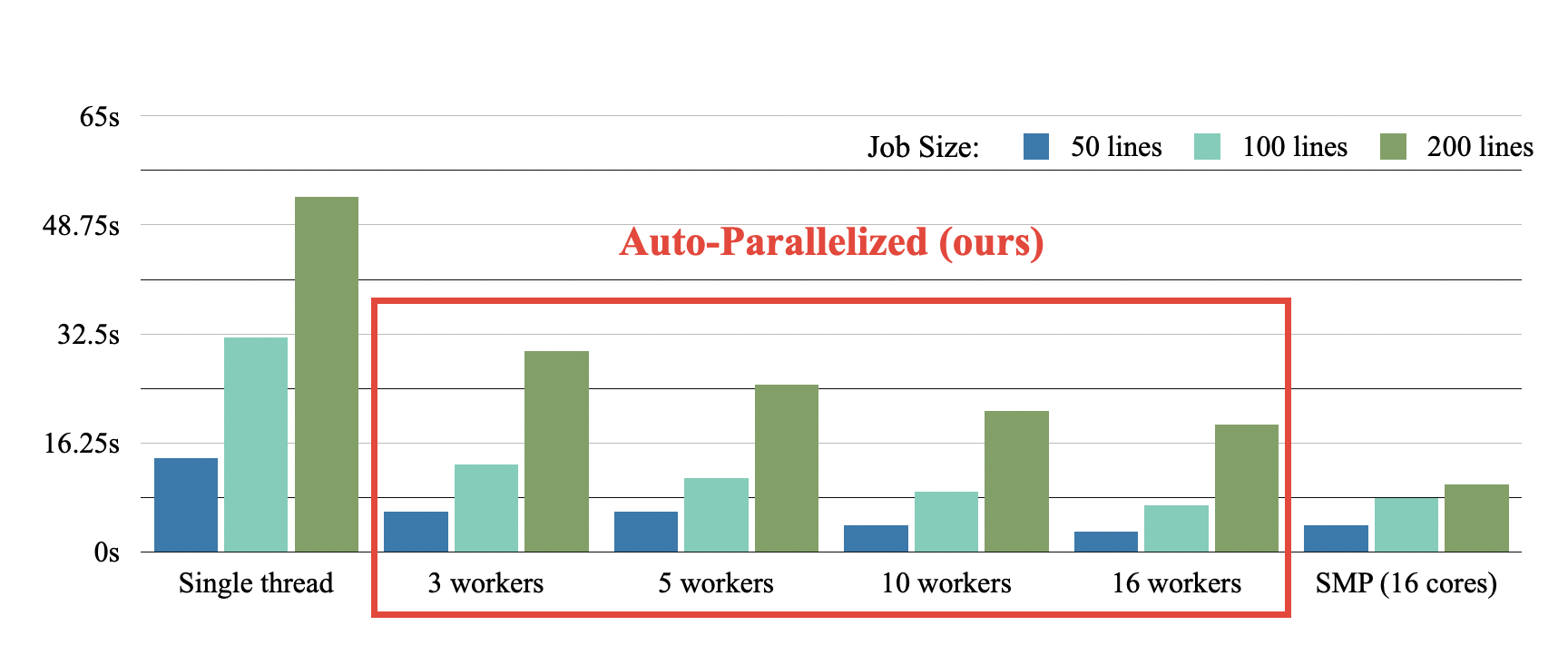}
    \caption{Benchmark results on large matrix multiplication tasks}
    \label{fig:fig}
\end{figure}

\begin{acks}
We would like to thank Chengrui Hou, Danyang Zhuo and Simon Peyton-Jones for their generous feedbacks.
\end{acks}

\bibliographystyle{ACM-Reference-Format}
\bibliography{refs}










\end{document}